\documentclass[11pt,preprint]{aastex}
\begin{document}
\def\etal{{\it et~al.}}
\def\asca{{\it ASCA}}
\def\rosat{{\it ROSAT}}
\title{The X-ray Line Emission from the Supernova Remnant W49B}
\author{Una Hwang (1,2), Robert Petre (1), John P. Hughes (3)}
\affil{(1) Laboratory for High Energy Astrophysics, NASA Goddard Space 
Flight Center, Greenbelt, MD 20771\\
(2) Department of Astronomy, University of Maryland, College Park, MD 20742\\
(3) Department of Physics and Astronomy, Rutgers University, 136 Frelinghuysen Road, Piscataway, NJ 08854-8019}

\begin{abstract}
The Galactic supernova remnant W49B has one of the most impressive
X-ray emission line spectra obtained with the {\it Advanced Satellite
for Cosmology and Astronomy (ASCA)}.  We use both plasma line
diagnostics and broadband model fits to show that the Si and S
emission lines require multiple spectral components.  The spectral
data do not necessarily require individual elements to be spatially
stratified, as suggested by earlier work, although when \asca\ line
images are considered, it is possible that Fe is stratified with
respect to Si and S.  Most of the X-ray emitting gas is from ejecta,
based on the element abundances required, but is surprisingly close to
being in collisional ionization equilibrium.  A high ionization age
implies a high internal density in a young remnant.  The fitted
emission measure for W49B indicates a minimum density of 2 cm$^{-3}$,
with the true density likely to be significantly higher.  W49B
probably had a Type Ia progenitor, based on the relative element
abundances, although a low-mass Type II progenitor is still possible.
We find persuasive evidence for Cr and possibly Mn emission in the
\asca\ spectrum---the first detection of these elements in X-rays from
a cosmic source.
\end{abstract}

\section{Introduction}

When the Galactic supernova remnant W49B (G43.3-0.2) was first
detected as an X-ray source by the {\it Einstein Observatory}, its
centrally-bright morphology raised speculations that its X-ray
emission is produced by nonthermal synchrotron processes (Pye et
al. 1984), as in the Crab.  The subsequent discovery of a prominent Fe
K emission line blend with {\it EXOSAT} showed that the X-ray emission
is thermal, and probably dominated by supernova ejecta (Smith et
al. 1985).  The {\it Advanced Satellite for Cosmology and Astronomy
(ASCA)} has since provided the highest quality X-ray spectral data yet
available for W49B, revealing a spectacular array of prominent
emission lines from the elements Si, S, Ar, Ca, and Fe.

Fujimoto \etal\ (1995) analyzed the intensity ratios of the Ly$\alpha$
and He$\alpha$ lines (n=2$\rightarrow$n=1 transitions in the H- and
He-like ions, respectively) using the \asca\ data, and concluded that
the elements Si, S, Ar, and Ca cannot have the same average ionization
age for a given temperature, with Si and S having the largest
discrepancies.  With the ionization age defined as the product $n_et$
of the ambient electron density $n_e$ and the time $t$ since the gas
was shock-heated (Gorenstein, Harnden \& Tucker 1974), this implies
that each element either has a different density or was shocked at a
different time.  This result is important, if true, since all the
elements Si, S, Ar, and Ca occupy essentially the same spatial zones
in supernova models, while Fe initially occupies a spatial zone
interior to these elements (e.g., Nomoto \etal\ 1997a, Thielemann,
Nomoto, \& Hashimoto 1990).  The \asca\ X-ray images of the remnant in
emission lines of Si, S, and Fe (with a low spatial resolution of FWHM
$>1'$) suggest that the Si and S emission is distributed outside more
centrally peaked continuum (4$-$6 keV) and Fe K emission.  These
results are interpreted as evidence for stratification of the
supernova ejecta, with the Fe ejecta lying interior to the Si and S
ejecta.

In this paper, we re-examine the \asca\ data for W49B, carrying out a
more complete and careful analysis of the X-ray line emission.  We
examine evidence for emission from less abundant elements, and
interpret the spectrum using both line intensity ratios and simple
broadband model fits.  We also present the previously unpublished
\rosat\ High Resolution Imager (HRI) data for W49B---essentially a Si
emission line image through the combined effect of the narrow
0.2$-$2.2 keV bandpass of the HRI and interstellar absorption of the
W49B spectrum at energies below $\sim$1 keV.  On the basis of our new
results, we suggest emission from multiple thermal components as an
alternative to the stratification scenario of Fujimoto \etal\ As
discussed later, however, our results do still allow for the possible
spatial separation of Fe from other elements.

After the submission of this paper, we were made aware of a similar
analysis of these data reported by Sun \& Wang (1999).  Our
independent analyses lead to similar general conclusions.

For completeness, we note that the remnant displays a sharply defined,
but incomplete radio shell of radius $\sim100''$, with high surface
brightness and very low polarization (Moffett \& Reynolds 1994, and
references therein).  Based on HI absorption measurements in the
radio, the distance to the remnant is 8 $\pm$ 2 kpc (Radhakrishnan et
al. 1972; adjusted for a Galactocentric radius of 8 kpc, following
Moffett \& Reynolds 1994).  To our knowledge, no optical emission has
been reported for this remnant, and it is likely that any such
emission is highly absorbed because of W49B's large distance through
the Galactic plane.  Absorption is virtually negligible for infrared
emission, however, and IRAS detected strong emission from the vicinity
of the remnant (Saken, Fesen, \& Shull 1992).  Though W49B is probably
a young remnant, its infrared colors are more consistent with an older
rather than a young remnant.  It is possible that that this is due in
part to source confusion in a crowded region.

\section{Data Reduction}

\asca\ (Tanaka \etal\ 1994) features two Solid-State Imaging
Spectrometers (SIS0 and SIS1) and two Gas Imaging Spectrometers (GIS2
and GIS3), each with a dedicated mirror.  The point-spread function
(PSF) of each of the X-ray mirrors has a narrow core of about $1'$
FWHM, and a half-power diameter of $3'$.  Each SIS is a square array
of four CCD chips, of which 1, 2, or all 4 may be exposed at a time
(1-, 2-, or 4-CCD mode).  The SIS provides moderate resolution
spectra between the energies 0.5 $-$ 10 keV, with the best instrument
performance in 1-CCD mode ($\Delta \rm{E} \sim 1/\sqrt{E} \sim$ 2\% at
6 keV at satellite launch in 1993).  Because we focus on the line
emission in this paper, and the GIS provides lower spectral and
spatial resolution data than does the SIS, we do not present the GIS
data.

There are ten \asca\ observations of W49B, all taken in 1993 during the
Performance Verification phase of the mission.  The source was
observed in a variety of instrument modes and detector positions
because it was used as a gain calibrator for the mission.  We examine
five of the observations in detail; we do not present the other five
because they have exposure times shorter than 3 ks, and in all but
one, the source is also either off or near the edge of the SIS field
of view.  Our observations are in different SIS instrument modes (1-,
2-, and 4-CCD), as noted in Table 1, with the source at different
positions on the detector.

Since the instrument performance and calibration for the early
observations are good, we use the REV2 cleaned data taken directly
from the HEASARC archive.  Interested readers may consult the \asca\
Data Reduction Guide (1997) for technical information on the data
processing procedures.  We focus on the 20 ks 2-CCD mode observation
because it has the longest exposure time at a single source position.
For fitting line intensities, we also consider a combined SIS spectrum
in which all the pulse-height spectra, response matrices, and
effective areas are weighted by the exposure time and averaged.  We
optimize the signal in the high energy spectrum by taking all the
spectra from a 3.2$'$ radius region centered on the source.  This
region contains roughly 75\% of the total flux from the source.

The background is taken from blank sky fields provided by the \asca\
Guest Observer Facility.  Since W49B is in the Galactic Ridge, we have
also verified that the results are not sensitive to the exact
background used.  We have compared the SIS0 2-CCD mode results using
the blank sky field with those using a local background taken away
from the source at the edges of the SIS, and find very good
consistency.  The only substantial differences are between the
best-fit line intensities of the weak Fe features beyond the strong Fe
K blend, but the 90\% confidence error ranges are in good agreement.
Roughly 80\% of the total SIS counts at energies above 1.5 keV come
from the source using either background subtraction.

The \rosat\ High Resolution Imager has a bandpass of 0.2$-$2.2 keV and
provides a spatial resolution of 5$''$ at the center of its 38$'$ wide
field of view.  A 34 ks observation (\rosat\ sequence rh500098n00)
centered on W49B was processed with software provided by Snowden \&
Kuntz (1998) to correct for vignetting, and to model and subtract the
particle background (Snowden 1998).  The image shown in Figure 1 is
smoothed using an adaptive filter with a varying spatial scale
corresponding to 50 counts per beam.  W49B is a weak \rosat\ HRI
source ($\sim$0.05 HRI counts/s) because the high column density
effectively cuts off the spectrum below 1 keV.  Compared to the {\it
Einstein}\ HRI, the \rosat\ HRI has higher sensitivity and lower
background, but a narrower bandpass.  The two bright spots that
dominate the {\it Einstein}\ HRI image (Pye \etal\ 1984, Seward 1990)
are clearly seen in the \rosat\ image, but the \rosat\ image also
shows the surrounding diffuse emission more clearly.  The diffuse
emission appears to be interior to the radio shell, and shows no sign
of limb-brightening.

\section{Spectral Results}

We discuss the \asca\ spectral results for W49B in this section,
treating the line intensities and plasma diagnostics first, followed
by the fitting of simple broadband spectral models.

\subsection{Line Spectrum}

W49B boasts one of the most impressive emission line spectra obtained
by \asca, featuring prominent He$\alpha$ blends (n=2$\rightarrow$n=1
in the He-like ion) and Ly$\alpha$ transitions (n=2$\rightarrow$n=1 in
the H-like ion) of Si, S, Ar, and Ca, clearly discernable higher level
transitions ($1s3p\rightarrow 1s^2\ [{\rm hereafter}\ 3p]$,
$1s4p\rightarrow 1s^2\ [{\rm hereafter}\ 4p]$), and a prominent Fe K
blend.  The line fluxes may be measured by modelling the continuum and
fitting narrow gaussian functions for each distinct emission feature.
Figure 2 shows the spectrum combining all the SIS data with the
best-fit line model (to be described below).

Line blending can be a significant issue when measuring line fluxes
with a moderate resolution spectrometer like the SIS.  Two important
instances of line blending that we address involve the Ly$\beta$ and
the $3p$ and $4p$ transitions.  At the \asca\ SIS resolution,
Ly$\beta$ lines are often blended with the He$\alpha$ emission of a
higher atomic weight element, and can have a non-negligible flux if
the corresponding Ly$\alpha$ lines are strong.  At the temperature of
the W49B continuum ($kT \sim$ 2 keV), the ratio of Ly$\beta$ to
Ly$\alpha$ intensity is between 0.11 and 0.14 for Si, S, and Ar in the
models of Raymond \& Smith (1977, hereafter RS).  For example, in
W49B, this means that S Ly$\beta$ is responsible for about 25\% of the
flux that would have been attributed to Ar He$\alpha$.  In our fits,
we therefore include the Ly$\beta$ lines of Si, S, and Ar at their
expected energies with an appropriately fixed intensity relative to
the Ly$\alpha$ line.  The $4p$ transition is also sometimes blended
with other lines, but fortunately, its intensity relative to the $3p$
transition does not change strongly with temperature and is constant
with ionization age.  We therefore model the the $4p$ line with an
intensity relative to $3p$ fixed at 0.6, as is appropriate for
temperatures between 1$-$2 keV in the RS models.  For Fe, we note that
$3p$ is blended with the He$\alpha$ transitions of Ni at the \asca\
resolution, and that its true flux is overestimated since we don't
model the Ni lines.

The energy scale of the W49B spectrum is established by its strong
Ly$\alpha$ lines, since they are effectively single transitions at
known energies.  The measured Ly$\alpha$ energies are in excellent
agreement with the expected values, with formal 90\% confidence errors
that are well within the nominal estimated 0.5\% accuracy of the SIS
gain; for Ca Ly$\alpha$, the formal error is comparable to the nominal
gain accuracy.  In our final fits, we fixed the energies of the
Ly$\alpha$ lines at their expected values and adjusted the overall
gain of the spectrum by 0.25\%.  The line energies of the He$\alpha$
blends are always freely fitted because they depend on the ionization
state of the gas: not only do the relative intensities within the
triplet of the He-like ion depend on the ionization state, but ions
less ionized than the He-like stage may also make a significant
contribution to this blend through lines with lower energies.  This
centroid is thus potentially a valuable diagnostic.

We fitted both the 2CCD-mode SIS0 spectrum and the spectrum combining
all the SIS data at energies between 0.6 and 10 keV with a model for
the continuum and the line features.  Lines of Si, S, Ar, Ca, and Fe
were included as described above and their fitted energies (for
He$\alpha$) and intensities are given for the combined SIS data in
Table 2 with their 90\% confidence ($\Delta\chi^2$ = 2.71) errors.
The continuum was modelled as two bremsstrahlung components, with
temperature about 1.7 keV for the dominant component, and fixed at 0.2
keV for the second component.  The inclusion of a second continuum
component gives a better fit, gives more consistent line diagnostics
results, and is consistent with the results of the broadband fits to
be described below.  The fitted column density is about $5 \times
10^{22}\ \rm{cm}^{-2}$ in the double continuum model, which is
significantly higher than the value of $3 \times 10^{22}\
\rm{cm}^{-2}$ when only one continuum component is modelled; the
Galactic column density measured in the radio is 1.8 $\times 10^{22}\
\rm{cm}^{-2}$ (Dickey \& Lockman 1990).  The X-ray measured absorption
column density is significantly higher than the radio value, but this
appears to be typical for neutral H column densities in excess of
several times $10^{20}$ cm$^{-2}$ because of the presence of
interstellar molecular gas (Arabadjis \& Bregman 1999).  The absolute
Si line intensities are therefore significantly different depending
how the continuum is modelled, but the Si line intensity ratios
discussed below are much less so.

\subsection{Plasma Diagnostics}

Measured line intensity ratios and He$\alpha$ centroid energies can
provide joint constraints on the emission-averaged temperature and
ionization age if the lines are chosen to eliminate or minimize other
dependences, such as element abundances or interstellar absorption.
The ionization age---defined as $n_e t$, where $n_e$ is the ambient
electron density, and $t$ is the time since the gas was
shock-heated---parameterizes the ionization state of the gas
(Gorenstein, Harnden \& Tucker 1974).  Gas that is suddenly heated by
the passage of the supernova shock wave is ionized slowly through
electron-ion collisions on a timescale $10^4/n_e$ yr, where $n_e$ is
in cm$^{-3}$.  This timescale may be comparable to, or larger than,
the known or deduced age of the remnant for a typical ambient density
of 0.1 cm$^{-3}$; most remnants are therefore expected not to be in
ionization equilibrium.  Since nonequilibrium ionization (NEI) affects
the ion population, and line intensities are directly proportional to
the population of the relevant ion, it can strongly affect X-ray line
intensities.  We calculate line emissivity ratios to compare with the
observations for a grid of temperatures and ionization ages, using the
updated code of Raymond \& Smith (1977, hereafter RS) for the X-ray
emission, and the matrix diagonalization code of Hughes \& Helfand
(1985) for the nonequilibrium ion fractions.

In Figure 3, we show the constraints on $kT$ and $nt$ based on 90\%
confidence limits for both the Ly$\alpha$ to He$\alpha$ and the
He$3p+4p$ to He$\alpha$ intensity ratios.  Separate panels show the
results for each of the elements Si, S, Ar, Ca, and Fe.  The observed
line intensity ratios are taken from the fit to the combined SIS data
with 90\% limits determined from the two-dimensional
$\Delta\chi^2$=4.61 contours.  These limits are consistent with, but
significantly tighter than, those from the 2-CCD mode SIS0 fits.  The
values of $kT$ and $nt$ allowed for Ar, Ca, and Fe are consistent with
each other near ionization equilibrium at kT = 2.2$-$2.7 keV, with the
exception that the Ca He$3p+4p$/He$\alpha$ ratio is marginally low.
Considering that we have not included systematic uncertainties in our
errors, the agreement between these three elements is very good, as
shown by the overlaid contours in the last panel of Figure 3.  The
temperature is somewhat higher than the temperature of 1.7 keV
inferred for the bremsstrahlung continuum.  For the elements Si and S,
however, the ratios are inconsistent with each other as well as with
the results for Ar, Ca, and Fe.  Since it is not possible for any
single spectral component to simultaneously reproduce the line ratios
for either Si or S, the emission from these elements must be more
complex.  For Si, this conclusion is supported by the Si He$\alpha$
line centroid.  Its constraints are inconsistent with those from the
ratio Si Ly$\alpha$/He$\alpha$, although a 0.5\% systematic error on
the line energy allows a small region of overlap near collisional
ionization equilibrium (CIE; see Figure 4).  The systematic errors on
the other centroids are too large to confirm the results of the S line
intensity diagnostics or to provide any new constraints for the other
elements.

If we assume that one spectral component has a temperature of 2.2 keV
and is at CIE, we may then infer parameters characterizing the second
component.  While the 2.2 keV CIE component gives the wrong Si
centroid, the $kT$ and $nt$ values that do give the observed Si
centroid also give low Ly$\alpha$ and He$3p+4p$ intensities relative
to the He$\alpha$ intensity.  Thus, the second component could
reproduce the Si He$\alpha$ centroid and provide nearly all of the Si
He$\alpha$ intensity, while contributing relatively little to
Ly$\alpha$ or He$3p+4p$.  Because of its low temperature, this
component contributes less to the S emission, and almost nothing to
the emission of the other elements.  The broadband fits described
below confirm that the single-temperature CIE model that accounts for
most of the W49B spectrum is deficient in Si He$\alpha$ flux.

Our results may be compared to the earlier results of Fujimoto et
al. (1995), who carried out a similar analysis for W49B using the same
data.  They use primarily the Ly$\alpha$ to He$\alpha$ line intensity
ratios for Si, S, Ar, and Ca and, as noted earlier, conclude that each
of these elements has a different ionization age for a given
temperature, with the Si and S parameters being the most disjoint.  We
are able to reproduce their results by modelling the spectrum with one
continuum component, and not including weaker lines in the model.
When we account for the Ly $\beta$ lines, however, we find very good
consistency between the parameters for Ar, Ca, and Fe.  We also show
that the Si and S emission requires multiple spectral components:
their line ratios (and the Si centroid) are strongly inconsistent with
a single temperature and ionization state.  We conclude that the
spectral data do not necessarily require stratification of the
individual elements, as Fujimoto et al. concluded, but that they do
reveal the spectral complexity of the emission from the elements Si
and S.

\subsection{Evidence for Emission from Cr and Mn}

As the spectral resolution and efficiency of X-ray spectrometers
continue to improve, it becomes feasible to search for emission from
elements other than the dozen most abundant ones that are now
routinely included in spectral models.  In Figure 5, the combined SIS
spectrum of W49B between the energies 5.0 and 6.4 keV shows line-like
features at energies of $\sim$ 5.7 and 6.1 keV.  There are no emission
lines in the RS model within the 90\% confidence errors of the
centroids, but these models do not include the elements Cr and Mn.
Emission lines of Cr and Mn do have energies near those of the
observed line features.  The forbidden and resonance transitions of
He-like Cr are at 5.655 and 5.682 keV, respectively, and those of Mn
at 6.151 and 6.181 keV; the Ly$\alpha$ transitions of Cr and Mn are at
5.917 and 6.424 keV, respectively.

The fluxes and centroids of these features were determined by modelling
each as a gaussian component along with a bremsstrahlung continuum for
the portion of the spectrum between energies of 4.5 and 6.4 keV.  The
lower energy feature, which we tentatively attribute to Cr, appears at
a line energy of 5.685$^{+0.020}_{-0.027}$ keV and has a flux of
3.0$^{+0.8}_{-1.1}\times 10^{-5} \rm{ph/cm}^2 \rm{/s}$, corresponding
to an equivalent width of about 90 eV.  For the other feature,
attributed tentatively to Mn, the line energy and flux are 6.172
$^{+0.047}_{-0.049}$ keV and 1.3 $^{+1.4}_{-0.6} 10^{-5}$ cm$^{-3}$
s$^{-1}$, or an equivalent width of 60 eV.  The reduction in $\chi^2$
from adding the ``Cr'' feature is more than 20 for 61 degrees of
freedom, while the reduction in adding ``Mn'' is an additional 10.  An
F-statistic of greater than 10 for 2 and 60 degrees of freedom gives a
probability greater than 99\% that these features are real.

There are no atomic data currently available for the calculation of
emissivities for either Cr or Mn.  To check if the strengths of the
observed features are consistent with our interpretation of them, we
have carried out the tests summarized in Figure 6.  We calculate the
intrinsic He$\alpha$ emissivity for Si (Z=14), S (Z=16), Ar (Z=18), Ca
(Z=20), Fe (Z=26), and Ni (Z=28) for the 2 keV CIE component without
including factors for the element abundances.  The crosses in the top
panel of the figure shows these calculated emissivities plotted
against atomic number, overlaid with a spline fit through the
individual points.  The calculated emissivities are then used with the
measured He$\alpha$ line fluxes to estimate the element abundances for
Si, S, Ar, Ca, Fe, and Ni (we use the Fe $3p+4p$ flux as an upper
limit for the Ni He$\alpha$ flux).  These abundances are plotted in
the bottom panel of the figure as crosses with error bars after
normalizing to the Ar abundance.  Ar was chosen for the normalization
as the lowest atomic number element for which most of the emission
comes from the CIE component.  We expect both Si and S to have
additional emission from a soft component; the high abundances of Si
and S (Si is off the scale of the plot) are spurious since their
He$\alpha$ emission arises in part from a cooler NEI component (see
the following section).  For the elements Cr (Z=24) and Mn (Z=25), we
interpolate the emissivities using the curve in the top panel of the
figure, and attribute all the measured flux at 5.7 and 6.1 keV to
the Cr and Mn He$\alpha$ blends, respectively.  The resulting
abundances are also plotted in the figure as crosses.  They are seen
to be consistent with the solar photospheric ratios (Anders \&
Grevesse 1989), which are plotted for each element as open circles in
the same panel.  As Cr, Mn, and Ni are the next most abundant elements
with K-shell emission lines at energies above $\sim$ 4 keV, they are
thus the atomic species most likely to be detected next.

\subsection{Broadband Spectral Fitting}

We turn next to fitting of the broadband \asca\ spectrum of W49B.  We
use collisional ionization equilibrium (CIE) models of RS,
nonequilibrium ionization (NEI) models for Sedov hydrodynamics of
Hamilton, Sarazin, \& Chevalier (1983; hereafter HSC) and single
temperature, single-ionization age models of RS with ion fractions
calculated according to Hughes \& Helfand (1985)---the model used for
the line diagnostics (\S3.2).  We expect that at least two spectral
components will be required, and first verify that no single spectral
component provides an adequate fit.  An RS model with temperature near
1.7 keV does best, but severely underpredicts the flux of the Si
He$\alpha$ blend and gives $\chi^2$ per degree of freedom greater than
3.  A comparable fit is obtained with an NEI model with a similar
temperature and an ionization age near 8$\times 10^{11}\ \rm{cm}^{-3}$
s.  Most of the X-ray emitting plasma in W49B is clearly at or near
CIE, as was first suggested by Smith \etal\ (1985) based on the Fe K
centroid and the Ly$\alpha$ to K$\alpha$ intensity ratio for the
measured continuum temperature of 1.8 keV.  The HSC Sedov models do
much worse than the other NEI models because the grid of models
available to us does not extend sufficiently close to CIE for the
relevant temperatures.

We obtain a satisfactory fit to the overall spectrum by combining a
CIE RS component with a low temperature NEI component.  Using a HSC
Sedov model for the NEI component gives the results shown in Table 3
and Figure 7.  The HSC component contributes significantly to the
emission lines of Si and S (providing nearly all of the Si He$\alpha$
blend) and to the surrounding continuum, but little at higher
energies.  As a Sedov model, it is itself intrinsically
multi-temperature.  If, instead, a single $kT$, single $nt$ model is
used  for the NEI component, it contributes primarily to the Si He
$\alpha$ blend, and very little to other lines.  Its contribution to
the spectrum falls off more quickly with energy than for the Sedov
component and its parameters are less well-constrained.  Models with
low temperatures between roughly 0.2 and 0.5 keV give good fits for a
range of ionization ages consistent with those required to give the
observed Si He$\alpha$ centroid.  If two NEI components are used to
model the spectrum, CIE is still favored for the dominant component,
with a $\Delta\chi^2=2.7$ range for $nt > 10^{12}$ cm$^{-3}$ s.

Although the absolute abundances of the elements are model-dependent,
it is worth examining the abundance results of our simple model fits.
To limit the number of free parameters in the models, we linked
element abundances together where possible.  The CIE component
contributes strongly to the line emission from Si, S, Ar, Ca, and Fe,
so the abundances of these elements are fitted freely for this
component.  The Mg and Ni abundances are also fitted because Mg and Fe
L emission are blended together at energies near 1.3 keV, with the Fe
L atomic data having known deficiencies (Liedahl et al. 1995), while
Ni and Fe emission are blended together at energies near 7.8 keV.  The
abundances of C, N, O, and Ne are fixed at their solar value---a
sufficient assumption since this spectrum is strongly attenuated by
absorption at the relevant energies below about 1 keV.  The data
cannot constrain a second full set of abundances for the NEI component
so the C, N, O, Ne, and Mg abundances are fixed at their solar value,
Si through Ca tied in their solar ratios, and Fe tied to Ni.  The
abundance results are summarized in Table 3.  The primary CIE
component requires abundances for Si, S, Ar, Ca, and Fe that are
comparable to each other and significantly enhanced above the solar
value.  The Mg abundance is formally zero, but has a large error.  The
second, NEI component suggests an enhanced Si abundance and a rather
low Fe abundance, but these abundances are nearly consistent when 
their errors are considered (see Table 3).

\section{Discussion}

The results of the broadband fits indicate that the X-ray emission
from W49B is dominated by its ejecta, and that the ejecta are at or
very near CIE.  Ejecta-dominated remnants are expected to be
dynamically young, so a high ionization age requires a high density.
We obtain a lower limit for the internal density of the X-ray emitting
gas of 2 cm$^{-3}$ by using the X-ray emission measure for the CIE
component and assuming that it fills a sphere with the radius of the
radio shell.  The same value was obtained by Smith \etal\ (1985) from
the EXOSAT X-ray data.  Moffett \& Reynolds (1994) also cite this
value as the minimum density required for significant Faraday
depolarization of the radio emission.  The high density required to
explain the X-ray emission thus also allows a possible explanation for
the unusually low polarization in the radio.
The actual density would need to be even higher since the X-ray
emission does not appear to fill the radio shell and is clearly
centrally concentrated.  The high density places constraints on the
properties of the progenitor.  Massive progenitors of type O through
B0 clear a surrounding radius of 15 pc during their main sequence
lifetime through their fast stellar winds and photoionizing radiation
(e.g., see Chevalier 1990).  These progenitors do not seem plausible
for W49B, which has a radius of only about 5 pc for a distance of 8
kpc, unless it is still interacting with dense circumstellar material
ejected by a slow wind if the progenitor underwent a red giant phase.

The element abundances are compared with the calculated abundances for
various supernova progenitors in Figure 8.  We focus on the relative
abundances because we find that they are more consistent among the
various models than are the absolute abundances.  The 90\% confidence
limits for the measured abundances of Mg, S, Ar, Ca, Fe, and Ni
relative to the abundance of Si in the CIE component are determined
from the two-dimensional $\Delta\chi^2$=4.6 contours.  The calculated
abundances for two Type Ia models---the standard W7 model and a
delayed detonation model (WDD2)---plus Type II models for progenitor
masses of 13 and 15 M$_\odot$, are all taken from Nomoto \etal\
(1997ab), and normalized to the Si abundance.  All abundances are
given relative to the solar photospheric abundances of Anders \&
Grevesse (1989).  As can be seen from the Figure, the 90\% confidence
limit for the Mg abundance relative to Si is consistent with all the
Type Ia and low mass Type II models that we considered.  More massive
Type II progenitors would produce too much Mg relative to Si to be
consistent with the data.  All the models underpredict Ar relative to
Si, but the Type II models also underpredict S and Ca relative to Si.
The WDD2 Type Ia model appears to be favored overall for the CIE
component, but is still far from being consistent with all the
observations.

Other possible nucleosynthesis diagnostics are the relative abundances
of less abundant elements, such as Ti, Cr, and Mn.  According to the
nucleosynthesis models used for Figure 8, the Ti abundance relative to
Mn and Cr is depressed relative to solar in Type Ia explosions, while
the three are comparable in low-mass Type II explosions.  We
demonstrated above that Cr and Mn are in roughly their solar ratio
relative to Ar.  Our 90\% confidence limits on the Ti He$\alpha$ flux
is $2\times 10^{-5}$ cm$^{-2}$ s$^{-1}$ at an energy $\sim$ 4.97 keV
(the He-like resonance and forbidden transitions of Ti are at 4.966
and 4.977 keV, and overlap the He $3p$ and $4p$ transitions of Ca at
the \asca\ resolution).  Using Figure 6, the 90\% upper limit for the
Ti abundance relative to Ar is about 5.7 dex, or a few times the solar
value.  Thus our upper limit does not distinguish between the models,
but future measurements of these quantities may improve enough to make
such diagnostics useful.  Foremost is the need for calculated
emissivities for these elements so that measured line fluxes can be
converted into reliable element abundances.  Currently, the codes that
calculate X-ray emission do not include the elements Ti, Cr, or Mn.

The total mass of the ejecta may formally be calculated by assuming a
geometry for the remnant.  The \rosat\ image is essentially a Si plus
continuum image, showing two bright spots plus a shelf of fainter
diffuse emission, each contributing roughly half of the image counts.
Since the Fe K and hard continuum images of Fujimoto et al. (1995) are
less extended than the Si and S images, we conclude that the soft
spectral component is more spatially extended and that the hard
component comes from the two bright spots.  If we assume that the CIE
component comes from the two bright lobes and model them as spheres of
diameter 0.5$'$, the total mass of the CIE component is 1.6 M$\odot$.
This mass is marginally consistent with the ejecta mass from a Type Ia
explosion, but is based on uncertain assumptions about the geometry
and the relative abundances of elements that do not exhibit strong
emission lines.  Our simple mass calculation therefore does not
unambiguously differentiate between Type Ia and Type II progenitors.
The calculated Si and Fe ejecta masses, at 0.02 and 0.04 M$\odot$
respectively, are lower than the Si and Fe masses for either Type Ia
or Type II explosions.  The calculated masses increase if the CIE
component is more extended than assumed, and also if the soft NEI
component in W49B is associated with ejecta.

Our broadband spectral fit results indicate that the Si and S lines
have a significant contribution from the NEI component, and that Si
He$\alpha$ in particular comes almost entirely from it.  Meanwhile,
the Ar, Ca, and Fe K emission comes almost entirely from the hotter
CIE component.  The \asca\ images of Fe K and Si from Fujimoto \etal\
show a distinct difference in morphology that can be explained
naturally if, as we have assumed earlier, the two spectral components
that we have identified in W49B have different spatial distributions:
a compact component for Fe K and a more extended component for Si.  We
then expect that the Si Ly$\alpha$ emission should be morphologically
similar to the Fe K emission (since both come from the CIE component),
while Si He$\alpha$ should be morphologically different (since it
comes from the other, cool NEI component).  Fujimoto \etal\ show that
the radial profiles for these two Si lines actually look similar.
Since the \asca\ PSF is not much smaller than the remnant itself, and
photon statistics are limited for an image of a single line feature,
more sensitive and higher spatial resolution observations with Chandra
and XMM will be necessary to resolve these discrepancies.  If the
distribution of the Si Ly$\alpha$ emission is truly similar to Si
He$\alpha$ and distinct from Fe K, our spectral results would require
both cool and hot extended Si ejecta, and compact hot Fe ejecta,
implying spatial segregation of the Si and Fe ejecta.

It is also possible that the soft spectral component represents a cool
blast wave, as suggested by Smith \etal\ (1985).  If so, our Sedov
parameters require the explosion to have been very weak (E $< 10^{50}$
ergs), and the remnant age to be about 4000 yr---conclusions that are
qualitatively similar to those reached by Smith \etal \ We cannot be
certain that the soft component is actually the blast wave, however.
The soft emission, while extended, is not limb-brightened, and does
not appear to fill the radio shell in currently available X-ray
images.  We also obtained better fits with enriched Si abundance than
with the element abundances in solar ratios, suggesting that the soft
NEI component may be from ejecta, though the models used are in need
of updated atomic physics.  In any case, forthcoming observations of
the distribution of the Si emission will shed much light on this
issue.  The blast wave may be too faint to have been detected yet.

\bigskip

The \asca\ X-ray observations have provided a wealth of new
information about the unusual supernova remnant W49B.  It has a rich
and complex spectrum including previously undetected emission lines of
the elements Cr and Mn.  The remnant shows evidence for spatial
differences between the X-ray emission of different elements that are
not yet understood: they may be due to ejecta plus the blast wave,
which has not yet been positively identified in X-rays, or to ejecta
of different densities and spatial distributions.  W49B bears the
strong imprint of its progenitor, in that its X-ray emission is still
dominated by its ejecta, but it is also clearly the product of its
complex environment.  It is an unusual example of an ejecta-dominated
remnant wherein much of the gas is very near collisional ionization
equilibrium, indicating that this young remnant has a very dense
environment---whether interstellar or circumstellar.  The current data
and models do not conclusively resolve whether this was a Type Ia or
Type II explosion, nor whether the soft spectral component is from a
cool blast wave or from ejecta, but have raised tantalizing puzzles
and hints of new discoveries that may be solved with future
observations.  These are fortunately soon forthcoming.

\acknowledgments
This work was based on archival X-ray data provided by the HEASARC at
NASA Goddard Space Flight Center.  We thank Steve Snowden for
providing software to process the \rosat\ data.  UH acknowledges
partial support through the NASA Astrophysics Data Program, and JPH
acknowledges support through NASA grants NAG 5-4794, NAG 5-4871, and
NAG 5-6419.

\clearpage
\begin{figure}
\plotone{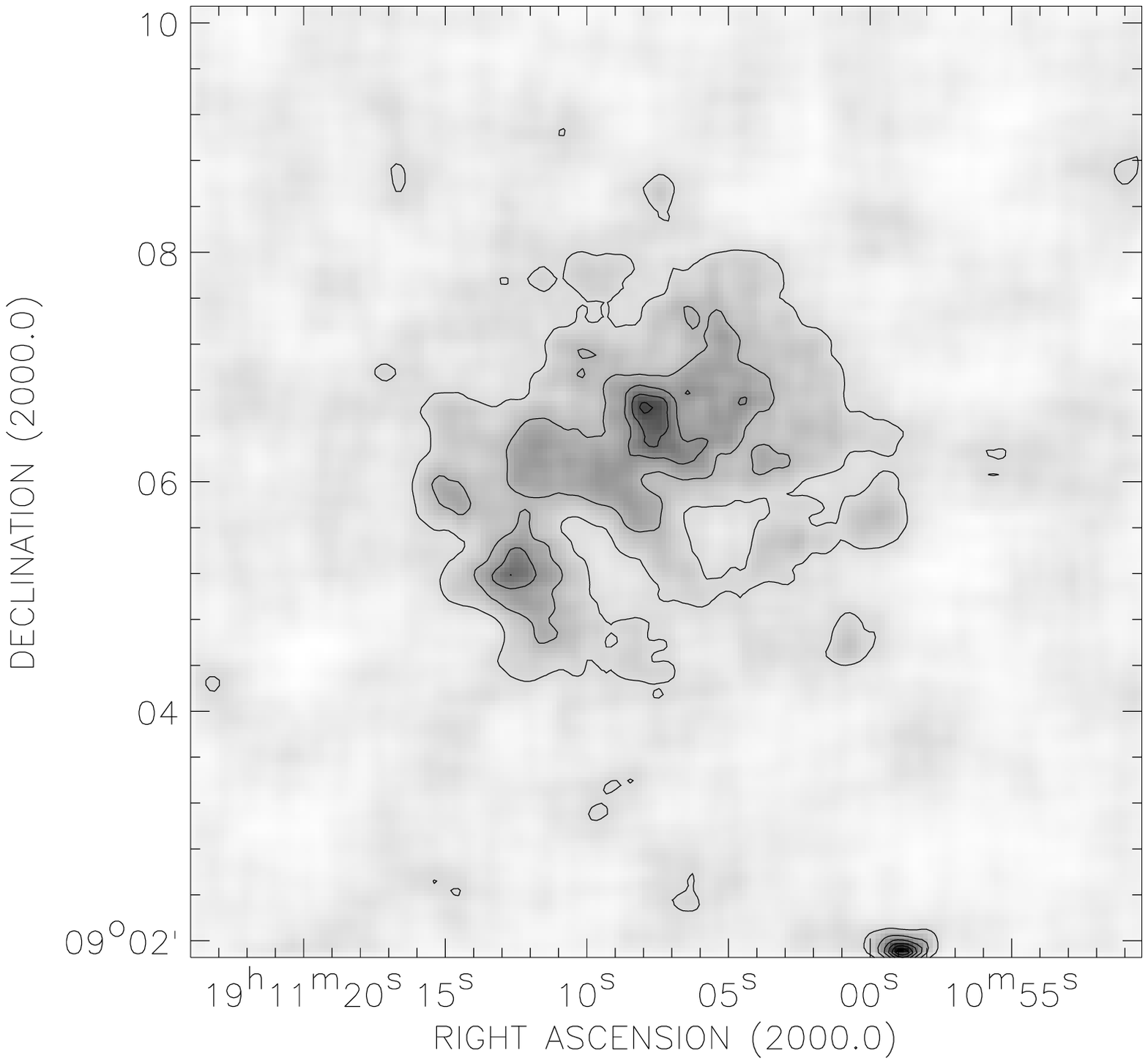}
\caption{Smoothed \rosat\ HRI image of W49B with contours overlaid
corresponding to 0.6, 1.2, 1.9, 2.6, 3.2, and 3.9 $\times 10^{-3}$
counts/s/arcmin$^2$.  The ROSAT image is dominated by the Si emission,
and shows two brighter spots surrounded by fainter diffuse emission
without signs of limb-brightening at the 100$''$ radius of the radio
shell.}
\end{figure}

\begin{figure}
\includegraphics[angle=-90,scale=0.80]{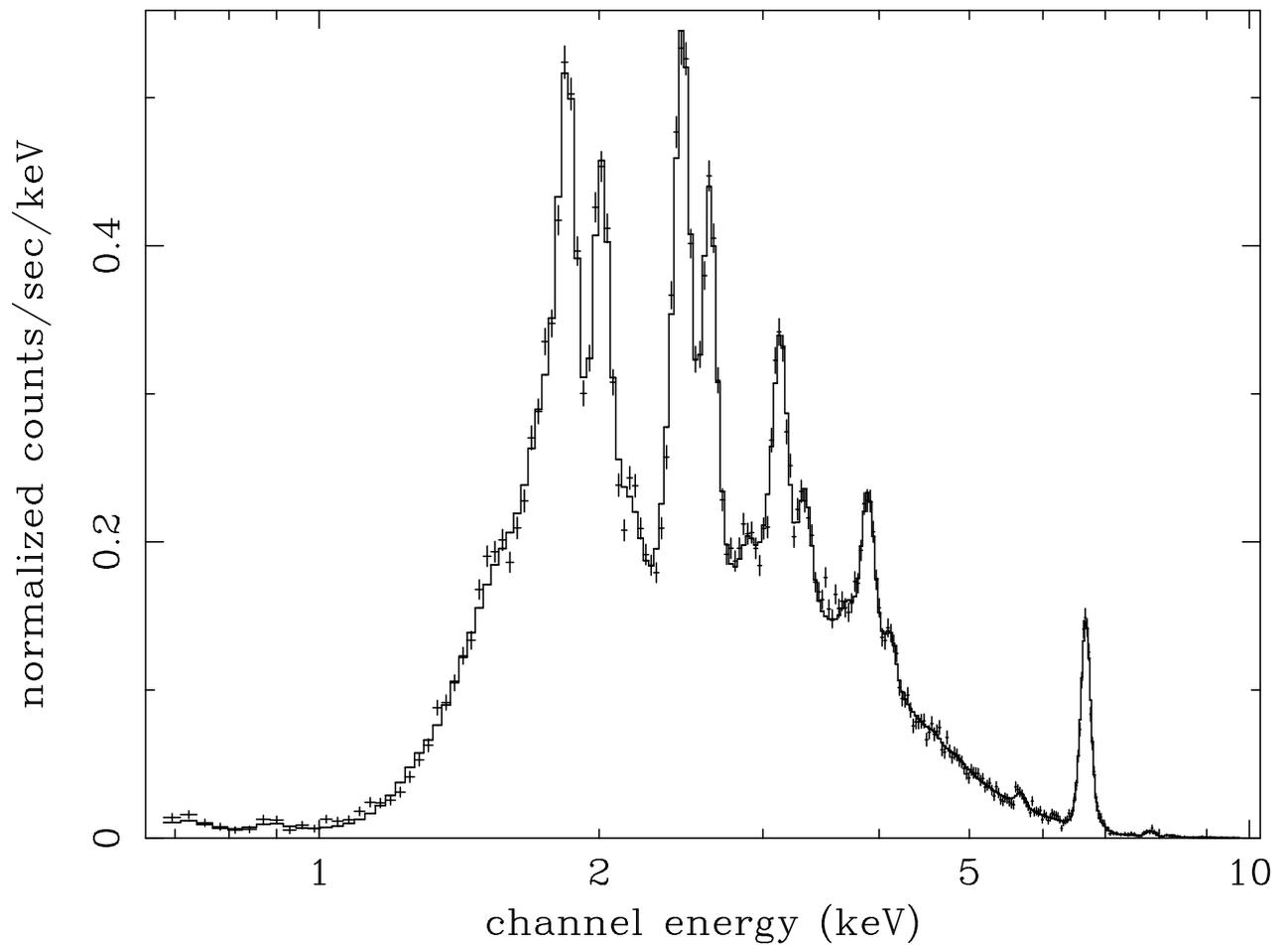}
\caption{Combined SIS data for W49B overlaid with the best fit {\it ad
hoc} spectrum for the continuum and lines.  This impressive spectrum
features prominent emission lines of Si, S, Ar, Ca, and Fe.}
\end{figure}

\begin{figure}
\epsscale{0.8}
\plotone{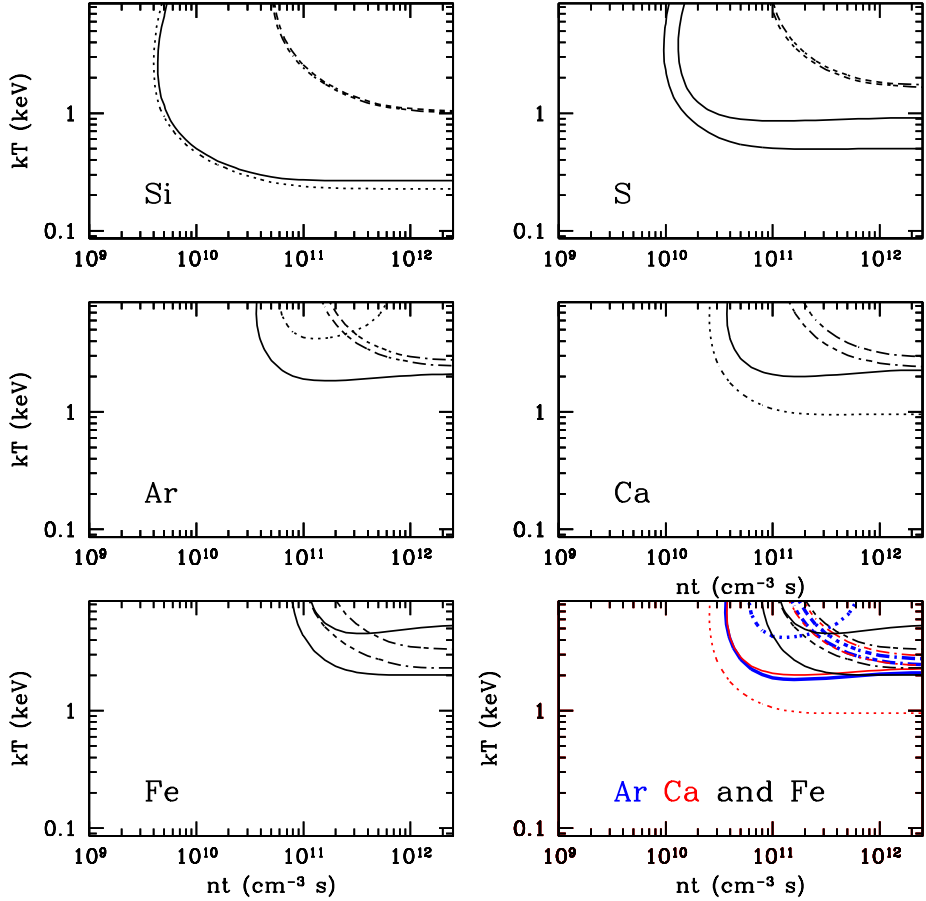}
\caption{Constraints on the emission-weighted average temperature
$kT$ (keV) and ionization age $nt$ (cm$^{-3}$ s) in W49B for the
elements Si, S, Ar, Ca, and Fe.  These are based on ratios of line
intensities from the fit to the combined SIS spectrum.  Each panel of
the figure shows the allowed region of parameter space for one
element, as indicated; the last panel shows the regions for Ar (blue),
Ca (red), and Fe (black) overlaid.  The region between the two
dot-dashed lines (in the upper right of each panel) is consistent with
the 90\% confidence lower and upper limits for the Ly$\alpha$ to
He$\alpha$ intensity ratio.  The solid lines show the region of
parameter space allowed by the limits for the He$3p+4p$ to He$\alpha$
ratio.  In the plots for Si and Ca, the lower limit for this ratio is
outside the boundaries of the plot, as is the upper limit for Ar.  For
these cases, the locus of points traced by the best-fit value is also
shown as a dotted line to clarify the region of parameter space
allowed by the data.  While the ratios of Ar, Ca, and Fe together
essentially allow temperatures of 2-2.5 keV with an ionization age at
or near CIE, the Si and S ratios are not consistent with these
parameters or with each other.}
\end{figure}

\begin{figure}
\plotone{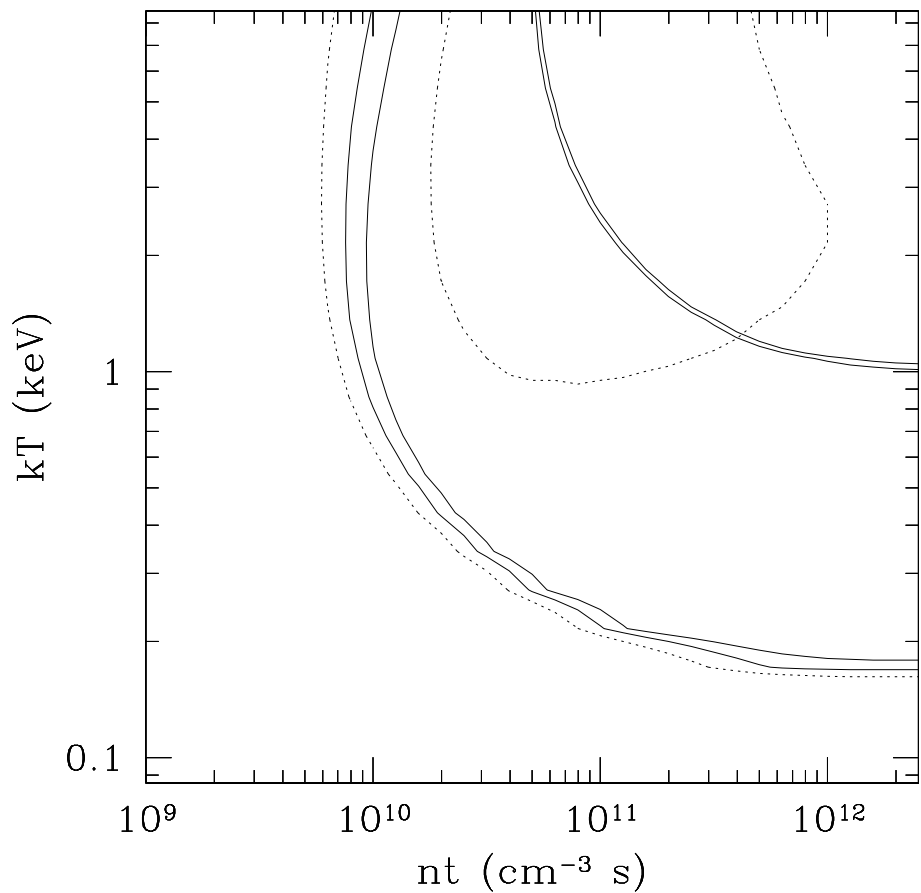}
\epsscale{0.6}
\caption{Constraints on the temperature and ionization age $nt$ of
the Si in W49B based on 90\% limits for the line ratio
Ly$\alpha$/He$\alpha$ (the two solid lines in the upper right) and the
He $\alpha$ centroid (the solid lines in the lower left).  The dotted
lines show 0.5\% systematic uncertainties in the centroid.  The
inconsistency between these diagnostics reinforces the conclusions
drawn from the previous Figure. }
\end{figure}

\begin{figure}
\includegraphics[angle=-90,scale=0.65]{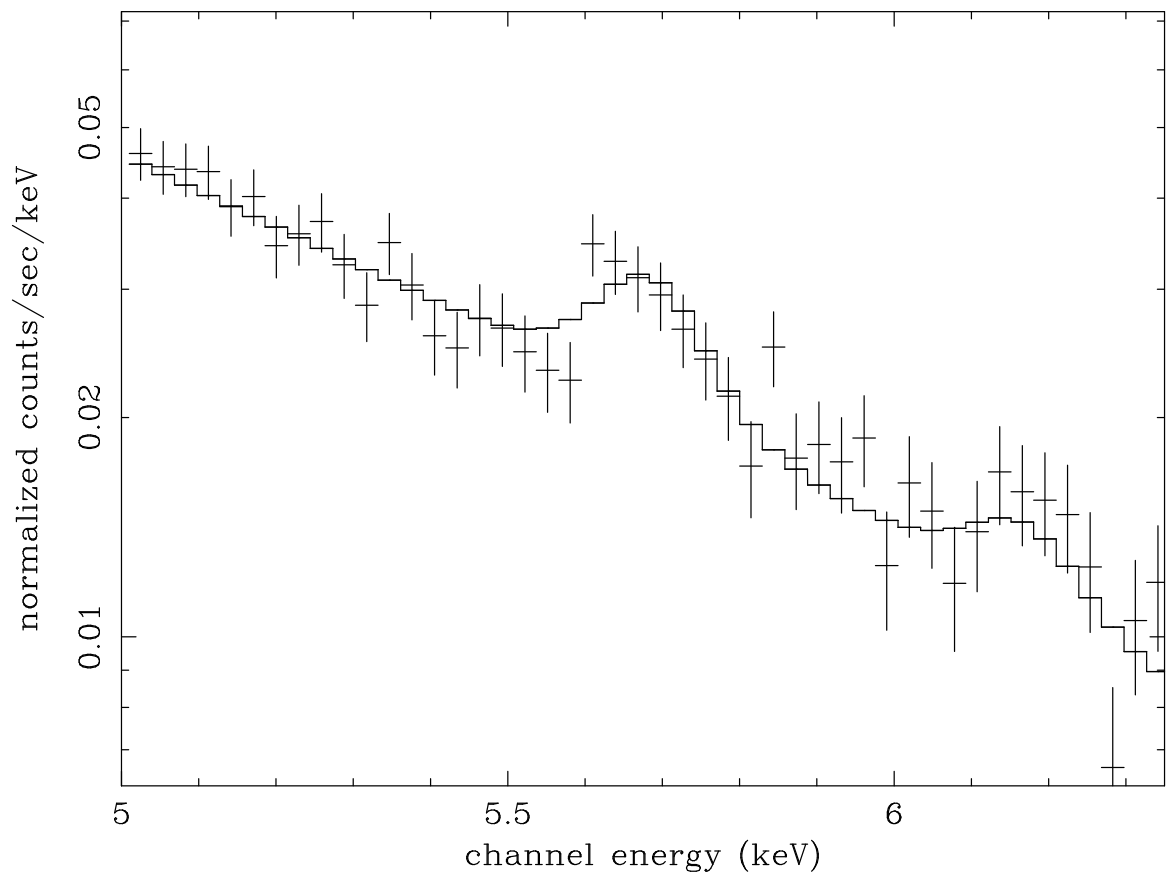}
\caption{A portion of the combined SIS data for W49B with a model for
a bremsstrahlung continuum fitted between 4.5 and 6.4 keV and two
gaussian line features overlaid.  The fitted line energies and
equivalent widths are 5.685 and 6.172 keV, and roughly 90 and 60 eV.
We identify the line features with Cr and Mn emission.}
\end{figure}

\begin{figure}
\epsscale{1.1}
\plotone{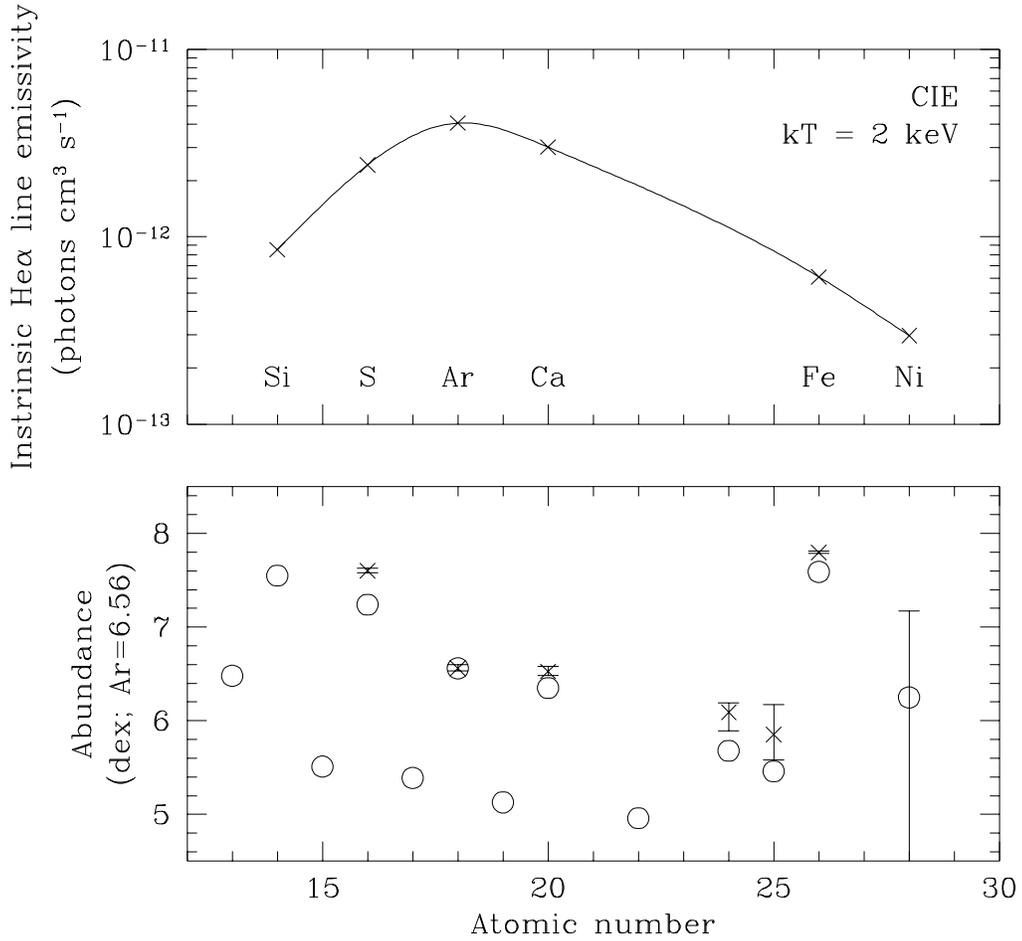}
\caption{Top panel: The crosses give intrinsic emissivities (not
including element abundance factors) for the He$\alpha$ blends of Si
(atomic weight Z=14), S (Z=16), Ar (Z=18), Ca (Z=20), Fe (Z=26), and
Ni (Z=28) calculated with the RS code for the 2 keV CIE component of
W49B.  The line is a spline fit to these points.  Bottom panel: The
crosses show abundances (normalized to Ar) calculated by using the
emissivities in the top panel and the measured He$\alpha$ line fluxes
in W49B.  See the text for more discussion.  For Cr (Z=24) and Mn
(Z=25), the emissivities were interpolated from the curve in the top
panel and all the flux at 5.7 and 6.1 keV, respectively, attributed to
the He$\alpha$ blend.  The solar photospheric abundance ratios of
Anders \& Grevesse (1989) are also shown as the circular points to
show that the implied abundances for Cr and Mn are consistent with a
solar ratio.  }
\end{figure}

\begin{figure}
\includegraphics[angle=-90,scale=0.65]{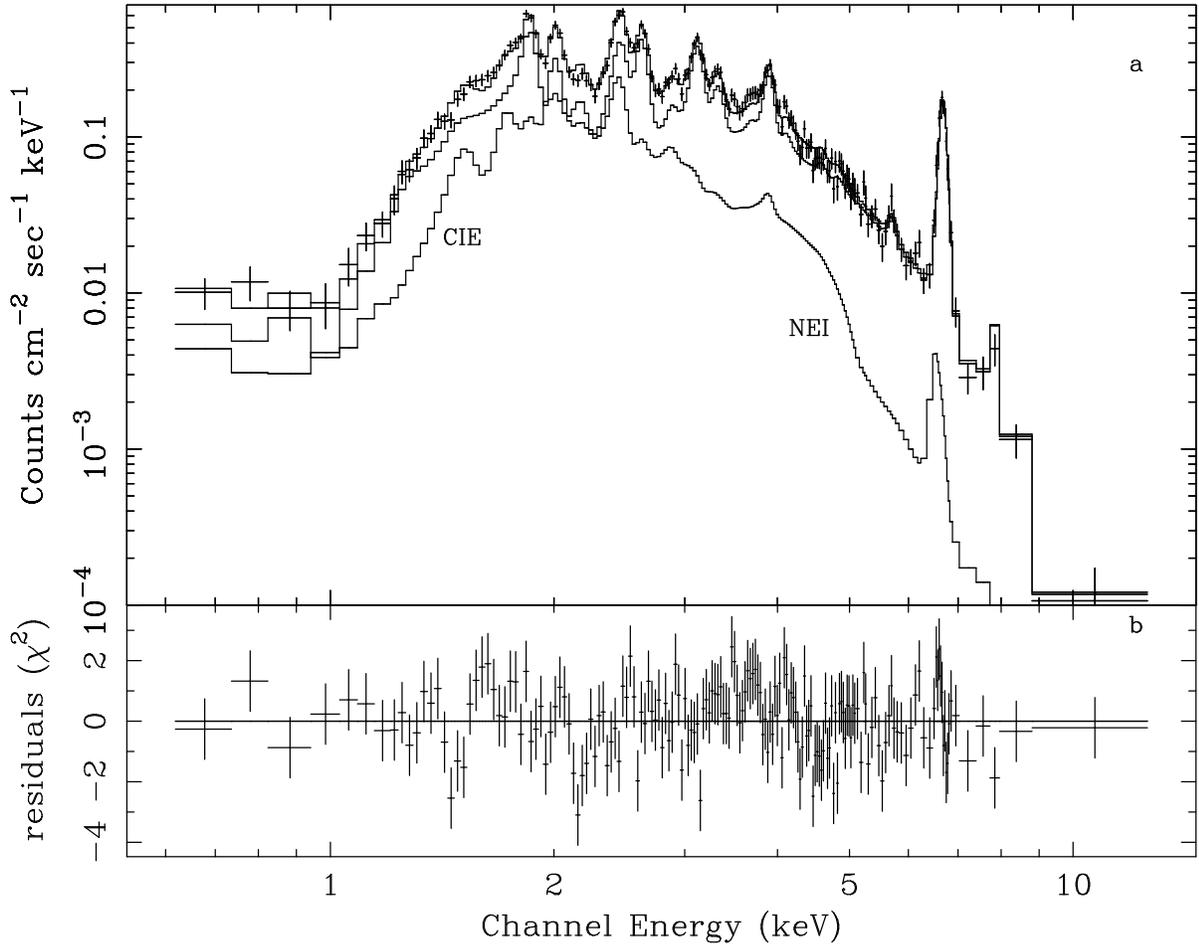}
\caption{The best-fit two-component model to the 2-CCD mode SIS0
spectrum of W49B, folded through the instrument response and overlaid
on the data.  The model includes a CIE component at temperature $kT$ =
2.0 keV and a Sedov NEI component with a shock temperature $kT$ = 0.15
keV and ionization age $nt$ = $5.3\times 10^{11} \rm{cm}^{-3}$s.  Each
component is also shown separately.  The soft NEI component falls off
more rapidly with energy, but contributes nearly all of the Si
He$\alpha$ flux, while the CIE component provides most of the flux in
the other emission lines.  The Cr feature at 5.708 keV has also been
included.  See the text for further discussion.}
\end{figure}

\begin{figure}
\epsscale{0.8}
\plotone{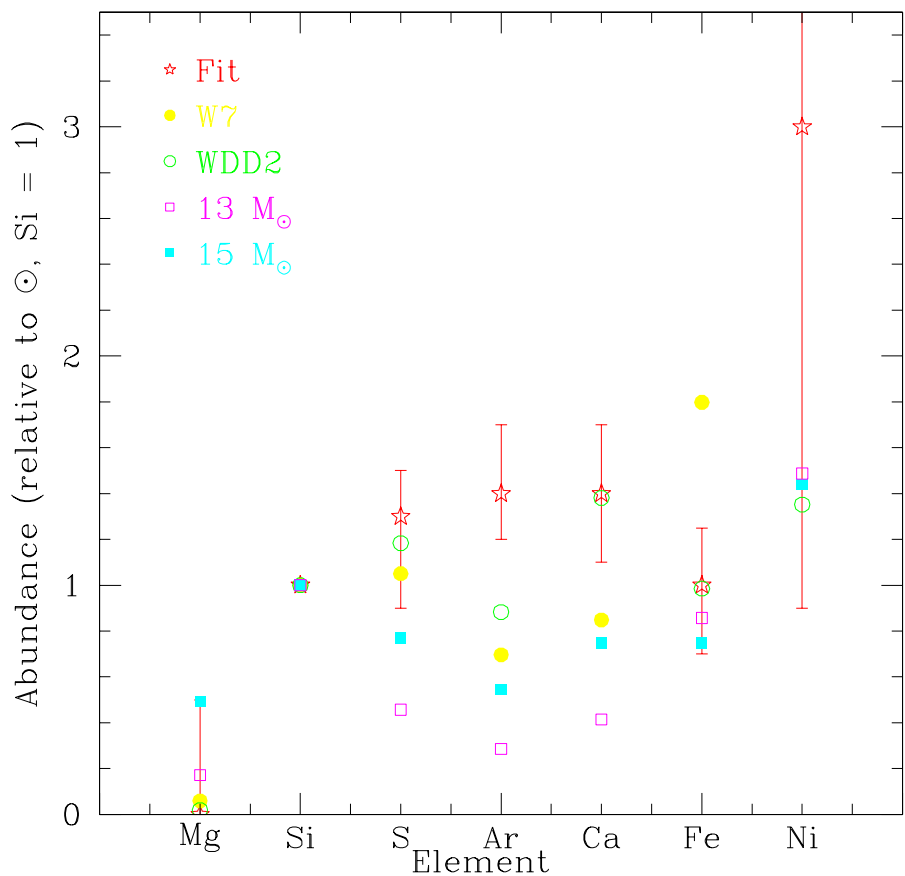}
\caption{The element abundances in the CIE component of the W49B
spectral model (open stars), normalized to the Si abundance relative
to the solar abundances of Anders \& Grevesse.  The error bars are for
90\% confidence ($\Delta\chi^2 = 4.61$ two-dimensional contours).
Also shown are the abundances in the Type Ia models W7(filled dots)
and WDD2 (open dots) and the Type II models for 15 (filled squares)
and 13 M$_\odot$ progenitors (open squares) of Nomoto et al. (1997ab).
The Type I models have more success overall, as the Type II models
tend to underpredict S, Ar, and Ca relative to Si.  All the models
appear to predict weaker Ar than is observed.  }
\end{figure}

\begin{deluxetable}{llcl}
\tablecaption{\asca\ Observations}
\tablehead{
\colhead{Date} & \colhead{Sequence} & \colhead{CCD Mode} & \colhead{SIS0 Exposure} \\
\colhead{} & \colhead{Number} & \colhead{} & \colhead{(ks)} \\
}
\startdata
1993 Apr 24 & 50005000 & 4 & 20.1 \\
            &          & 2 & 21.8 \\
1993 Oct 16 & 50005010 & 4 & 9.2  \\
1993 Oct 17 & 50005020 & 4 & 13.7 \\
1993 Nov 3  & 10020000 & 1 & 11.1 \\
1993 Nov 3  & 10020010 & 1 & 17.3 \\
  &  &  & 93.2 total \\
\enddata
\end{deluxetable}

\begin{deluxetable} {llll}
\scriptsize
\tablecaption{Line Fluxes (Combined SIS)}
\tablehead{
\colhead{Line} & \colhead{Fitted Energy} & \colhead{Unabsorbed Flux} \\ 
\colhead{ } & \colhead{(keV)} & \colhead{(10$^{-3}$ ph/cm$^2$/s)}\\
}
\startdata
Si He $\alpha$ & 1.848 (1.846$-$1.850)  & 6.7 (6.5$-$6.9) \\
Si Ly $\alpha$ & 2.006        & 3.2 (3.0$-$3.4) \\
Si He $3p+4p$  & 2.185, 2.294 & 0.15 (0.05$-$0.22) \\
S  He $\alpha$ & 2.455 (2.453$-$2.457)  & 2.1 (2.0$-$2.25) \\
S  Ly $\alpha$ & 2.623        & 1.3 (1.2$-$1.35) \\
S  He $3p+4p$  & 2.884, 3.033 & 0.16 (0.11$-$0.20) \\
Ar He $\alpha$ & 3.136 (3.131$-$3.141)  & 0.32 (0.30$-$0.35) \\
Ar Ly $\alpha$ & 3.323        & 0.19 (0.17$-$0.21) \\
Ar He $3p+4p$  & 3.685, 3.875 & 0.067 (0.043$-$0.090) \\
Ca He $\alpha$ & 3.880 (3.874$-$3.887)  & 0.22 (0.20$-$0.25) \\
Ca Ly $\alpha$ & 4.105        & 0.069 (0.057$-$0.080) \\
Ca He $3p+4p$  & 4.582, 4.818 & 0.011 ($<$ 0.023) \\
Fe K $\alpha$  & 6.658 (6.656$-$6.661)  & 0.83 (0.81$-$0.86) \\
Fe Ly $\alpha$ & 6.965        & 0.024 (0.013 $-$ 0.037) \\
Fe He $3p+4p$  & 7.798, 8.216 & 0.074 (0.048$-$0.096) \\
\enddata
\end{deluxetable}

\begin{deluxetable}{ll}
\tablecaption{Broadband Fit (2-CCD Mode SIS0)}
\tablehead{
}
\startdata
$\chi^2$, DOF    & 219.7, 163 \\
$N_H\ (10^{22}\ \rm{cm}^{-2})$ & 5.0 (4.8$-$5.3)\\
\multicolumn{2}{c}{CIE component} \\
$kT$ (keV)       & 2.0 (1.9$-$2.1) \\
$EM$ ($n_e n_H V/4\pi d^2$ in cm$^{-5}$) & $4.9 \times 10^{12}$ \\
Unabsorbed Flux (0.5-10. keV in $10^{-10}$ ergs/cm$^2$/s) & 1.5 \\
Mg ($\odot$)     & 0 ($<$1.7) \\
Si ($\odot$)     & 5.0 (3.7$-$7.0) \\
S  ($\odot$)     & 6.4 (4.9$-$8.5) \\
Ar ($\odot$)     & 7.1 (5.4$-$9.5) \\
Ca ($\odot$)     & 7.0 (5.5$-$9.0) \\
Fe ($\odot$)     & 4.8 (4.0$-$5.8) \\
Ni ($\odot$)     & 14.9 (7.4$-$23) \\
\multicolumn{2}{c}{HSC NEI component} \\
$<kT>$ (keV)     & 0.15   \\
$<nt>$ (10$^{11} \rm{cm}^{-3}$ s) & 5.3 \\
Unabsorbed Flux (0.5-10. keV in $10^{-10}$ ergs/cm$^2$/s) & 90\\
Si ($\odot$)     & 2.8 (2.1$-$3.9) \\
Fe ($\odot$)     & 0.9 (0.2$-$1.6) \\ 
\enddata
\end{deluxetable}

\end{document}